\documentclass[11pt]{article}
\pdfoutput=1
\usepackage[utf8]{inputenc}
\usepackage{graphicx}
\usepackage{amsmath}
\usepackage{amsfonts}
\usepackage{amssymb}
\usepackage{cite}

\usepackage[letterpaper]{geometry}
\title{Could Bert, Ernie and Big Bird be 13 billion years old? \\
\vspace{2mm}
\Large The Inverse Fermi scattering ($p\nu\rightarrow p'\nu'$) as\\ acceleration of cosmological neutrinos
}
\author{Guido Barbiellini}
\date{\today}

\begin{document}

\maketitle

Recently the IceCube Collaboration reported the neutrino number of events relative to 2078 days of data taking~\cite{b1}.

The IceCube Collaboration also presented the 3 highest energy events ($E_{\nu} \geq 1$~PeV) named by the collaboration as Bert (1.0\,PeV), Ernie (1.1\,PeV) and Big Bird (2.2\,PeV). The energy range covered by the 3 events does not contain background contamination~\cite{b1}. The 3 events correspond to the ID 4, 20 and 35 of Suppl.Tables in~\cite{b2}. The number of events is an order of magnitude larger than the one expected from UHE protons ($2\cdot 10^{19}$\,eV) interacting on CMB (3\,K now) photons via the reaction:
\begin{equation}
\label{reac}
\gamma~p \rightarrow N~\pi^{\pm} \quad \Rightarrow \quad\pi \rightarrow \mu + \nu_{\mu}\,.
\end{equation}

Moreover the average energy of the 3 HE $\nu$ detected by IceCube is more than one order of magnitude lower than that expected by reaction \eqref{reac}. Since the lowest proton energy for the photon-$\pi$ production is $2\cdot 10^{19}$\,eV (energy threshold for $\pi$ production on CMB$\gamma$) then  $E_{\nu} \sim \left( \frac{1}{20} E_p \right)\sim 10^{18}$\,eV~\cite{b3}.

In the following note, based on the 3 HE $\nu$, is considered the hypothesis that the Cosmological $\nu$ are boosted by the first Cosmic Rays (protons) born in the Universe through the Inverse Fermi scattering:
\begin{equation}
\label{IFs}
p~\nu \rightarrow p'~\nu'\,.
\end{equation}

The reaction \eqref{IFs} is the Weak scattering analogous to the electromagnetic one, namely the Inverse Compton (IC).
The $\nu$ boosted by the CR proton get $\sim \frac{1}{2000} E_p$, in the present $\nu$ and proton energy range. 

The reaction \eqref{IFs} is an elastic scattering. Since the initial energy of the $\nu$ is known from the history of the Universe ($1.7\cdot 10^{-4}$\,eV now, \cite{b4}) and the one of the boosted $\nu$ is measured by the IceCube detector. It is possible to compute the $\gamma^2$ of the boost:
\begin{equation}
\label{gamma}
\gamma^2 = \frac{E_{\nu}^{(m)}(z=0)}{E_{\nu}^{(\rm cos)}(z=0)}\ ,
\end{equation}
where $E_{\nu}^{(m)}$ is the measured one, and $E_{\nu}^{\rm (cos)}$ is the cosmological one.

Taking as $E_{\nu}^m$ the average of the 3 HE $\nu$, $\langle E_{\nu}^m \rangle = 1.4$\,PeV: 
\begin{equation}
\label{four}
\gamma^2 = 0.83~10^{19}\,;\quad \gamma = 2.9~10^{9} \,;\quad E_p = \gamma m_p = 2700~10^{15}\,{\rm eV} \sim 
2.7~10^{18} {\rm eV}\,.
\end{equation}

It is worth noting that the $E_p$, derived from eq.~\eqref{four} is one order of magnitude lower that the $E_p^{\rm th}$ for the charged $\pi$ production on CMB, from eq~\eqref{reac}.

The kinematic of the Inverse Fermi (IF) process is simple and it is the same of the IC. In the following the approximate expectation of the number of events is computed with a parametrization of the redshift evolution of the proton flux. The final formula is:
\begin{equation}
\label{finaleq}
N_{\nu} = K \int_{1}^{1+z_{\rm d}}\!\!\!\! \frac{(1+z)^{2} \cdot f_{\rm csfr}^{\frac{3}{2}}}{\sqrt{\Omega_m(1+z)^3 + \Omega_{\Lambda}}}\,\, d(1+z) + \int_{1 +z_{\rm d}}^{1+z_{\rm m}}\!\!\!\! \frac{(1+z)^{2} \cdot f_{\rm csfr}^{\frac{3}{2}}}{\sqrt{\Omega_m(1+z)^3 + \Omega_{\Lambda}}}\,\, d(1+z), 
\end{equation}
where $K = Q_0 \sigma_W \rho_0 c H_0^{-1} (A_{\rm eff}/A)$, $f_{\rm csfr} = f_{\rm csfr}(z)/ f_{\rm csfr}(0)$ and in particular $f_{\rm csfr}(z)$ is the star formation rate at the redshift $z$ and $f_{\rm csfr}(0)$ it's value at $z=0$.
The cosmic-ray energy density evolution is assumed proportional to the $f_{\rm csfr}$ to the power $3/2$ over all the $z$ values~\cite{b5}. The $f_{\rm csfr}$ is measured from $z=0$ to $z\sim8$ and the first integral of formula~\eqref{finaleq} is computed assuming the experimental values of $f_{\rm csfr}$ so at $z_{d}\sim8$. The second integral, from $z_{d}\sim8$ to $z_{max}=20$, is computed assuming a mean value of $f_{\rm csfr}(14)\sim 15 \, \rm M_{\odot} y^{-1}Mpc^{-3}$ to match the hypothesis that the three PeV neutrinos are coming from cosmological $\nu$ boosted by the first accelerated high energy protons in the universe.

$Q_0$ is the proton flux at $z=0$ integrated over $t = 6$\,years of time, a surface of $A=1$\,km$^2$ and a solid angle of $\Delta \Omega \sim 4\pi$ at $E_p^0 = 2.7~10^{18}(z+1)^{-1}$~eV and  $\Delta E_p^0 = E_p^0$; $\sigma_W~\!=~\sigma (\nu p~\rightarrow~\nu' p')$ at 1\,GeV $= 10^{-38}$~cm$^2$; $\rho _0$ is the CNB density at $z = 0$; $c$ is the speed of light; $H_0$ is the Hubble constant.

The $\nu$ target density increases as $(1 + z)^3$ for any $z$ value, $\rho(z) = \rho_0 (1+z)^3$. The existence of Cosmic ray proton requires the presence in the Universe of stars and strong star explosions and this condition imposes a maximum value of z, compatible with the Universe evolution, for the existence of the IF.\,$A_{\rm eff}$ is the IceCube effective area for $\nu$ of $\sim 1$\,PeV; $A_{\rm eff} = 20$\,m$^2$ (from~\cite{b56}) so $A_{\rm eff}/A \sim 2\cdot10^{-5}$.

\section*{Conclusions}
The three highest energy $\nu$ detected by IceCube over 6 years of exposure could be generated by cosmological $\nu$ boosted by proton scattering if the cosmic-ray energy density at the time of the formation of the first stars (POP III) is $\sim 15 \rm M_{\odot} y^{-1}Mpc^{-3}$. The VHE neutrinos from POP III stars has been studied in~\cite{b6}.

\section*{Acknowledgements}
I am very grateful to all the colleagues and friends that has been interested to listen to this innovative interpretation of the origin of the highest energy neutrinos ever detected. 

In particular the High Energy Gamma Astrophysics group lead by Francesco Longo devoted some time to discuss the physics of the IF process. I would like to sincerely thank Thomas Gasparetto and Michele Palatiello for critically reading and editing the manuscript resulting in stimulating relevant discussions. Paolo Lipari mentioned to me that investigation on this direction was not present in literature and that was worthwhile to start with numerical evaluation.

I am also glad to thank Dr. Guo-Yuan Huang for a critical reading founding an error in the adiabatic universe expansion formula in the previous version.

\end{document}